\documentclass[epj]{svjour}
\usepackage{graphics}
\usepackage{epstopdf}
\usepackage{comment}
\usepackage{newtxtext,newtxmath,amsmath}
\usepackage[usenames,dvipsnames]{xcolor}
\usepackage{overpic}
\usepackage{placeins}
\begin{document}
\title{Parametric investigation of laser interaction with uniform and nanostructured near-critical plasmas}
\author{L. Fedeli\inst{1}\thanks{\emph{e-mail:} luca.fedeli@polimi.it}, A. Formenti\inst{1}, C.E. Bottani\inst{1} \and 
M. Passoni\inst{1}
}                     
\offprints{}          
\institute{\inst{1} Politecnico di Milano, Italy}
\date{Received: date / Revised version: date}
%
\abstract{Laser interaction with uniform and nanostructured near-critical plasmas has been investigated by means of 2D particle-in-cell simulations. The effect of a nanostructure (modeled as a collection of solid-density nanospheres) on energy absorption and radiative losses has been assessed in a wide range of laser intensities (normalized amplitude $a_0 = 1 - 135$) and average densities of the target (electron density $n_e = 1 - 9 n_c$, where $n_c$ is the critical electron density). The nanostructure was found to affect mainly the conversion efficiency of laser energy into ion kinetic energy and radiative losses for the highest simulated intensities.
}
%
\authorrunning{L. Fedeli et al.}
\maketitle
\section{Introduction}
\label{intro}
Near-Critical density Plasmas (NCPs) are a longstanding research topic in the laser-plasma interaction community \cite{FreidbergPRL1972,ReganPoP1999}.\\
NCPs are characterized by an electron density close to the transparency threshold.
For a given laser wavelength $\lambda$, the critical electron density $n_c$ is defined by the relation $\omega_p(n_c) = 2 \pi c/ \lambda$, where $\omega_p(n_e) = \sqrt{ \frac{4\pi n_e e^2}{m_e}}$ is the plasma frequency ($n_e$ is the electron density, $e$ is the elementary charge and $m_e$ is the electron rest mass). For sufficiently low laser intensities,  if $n_e \ll n_c$ the plasma is essentially transparent and a laser pulse can propagate for long distances, while if $n_e \gg n_c$ the plasma is opaque and an impinging laser pulse is reflected back, usually with limited absorption. The intermediate regime ($n_e \approx n_c$) is characterized by a strong coupling between the pulse and the plasma (e.g. via the excitation of bulk plasmons), which leads to a variety of physical effects and to a strong laser absorption.\\
In this work we are particularly interested in NCPs irradiated at relativistic laser intensities \cite{AskaryanJETP1994,PukhovPRL1996,BorghesiPRL1997} (i.e. using laser pulses with normalized amplitude $a_0= \frac{e A_0}{mc^2}$ $> 1$,  where $A_0$ is the  peak amplitude of the vector potential of the laser pulse). Taking relativistic effects into account  \cite{UmstadterScience1996}, the transparency threshold should be increased to $n_c^{rel} \approx \langle\gamma\rangle n_c$, where $\langle\gamma\rangle$ is the average Lorentz factor of the plasma electrons. For linearly polarized laser pulses $\langle\gamma\rangle \approx \sqrt{1 + a_0^2 / 2}$, while for circular polarization $\langle\gamma\rangle \approx \sqrt{1 + a_0^2}$ \cite{SpranglePRL1990}. Thus, analogously to what was done in \cite{RobinsonPPCF2011} for circularly polarized laser pulses, the near-critical density regime for relativistic laser intensities and linear polarization could be conveniently defined as:
\begin{equation} \label{eq:dens}
0.1 n_c < n_e < \sqrt{1 + \frac{a_0^2}{2}} n_c
\end{equation}
Relativistic laser interaction with NCPs is characterized by a rich physics and has been investigated for a wide variety of applications: laser-driven particle sources (electrons  \cite{GoersPRL2015}, positrons  \cite{zhuNatComm2016} and especially ions\cite{YogoPRE2008,WillingalePRL2009,NakamuraPRL2010,PRESgattoni2012,PassoniPPCF2014,BinPRL2015,PrencipePPCF2016,PassoniPRAB2016,BrantovPRL2016}),laboratory astrophysics (the proposed schemes to observe collision-less shock acceleration of ions rely on slightly overcritical plasmas  \cite{HaberbergerNatPhys2012,FiuzaPoP2013}), advanced strategies for Inertial Confinement Fusion   \cite{KodamaNature2001,WillingalePoP2011} and ultra-intense, ultra-short gamma sources \cite{NakamuraPRL2012,BradyPPCF2013,WangPoP2015,StarkPRL2016,ChangPoP2017}. Some recent results include enhanced high-order harmonic emission in NCPs   \cite{BlackburnARXIV2017} .\\
Despite the wide interest of the topic, experiments involving controlled NCPs are usually challenging from the targetry point of view.  Laser systems able to attain relativistic intensities  \cite{DansonHPL2015} are usually Titanium:Sapphire lasers ($\lambda \approx 0.8 \mu$m) or Neodimium-based lasers ($\lambda \approx 1 \mu$m). Thus the critical density corresponds to an electron density $n_c $ $\approx 10^{21} \textrm{cm}^{-3}$, which means 2-3 orders of magnitude lower then the typical solid density (e.g. for plastic $n_e \approx 200-300 n_c$, while for gold $ n_e> 10^3 n_c$). According to eq.\ref{eq:dens} cryogenic solid hydrogen targets  \cite{MargaronePRX2016}($n_e \approx 30 n_c$) could allow to reach the near-critical regime for PetaWatt-class laser systems  \cite{DansonHPL2015} able to attain $a_0 \gtrsim 30 $. NCPs with $n_e < n_c$ can be obtained with gas-jets  \cite{SyllaRevSciInst2012}, however this density regime is particularly challenging, since high-density cryogenic gas-jets are required. Another possible strategy consists in using exploding solid targets, pre-heated with another laser \cite{YogoPRE2008} or by the pulse pedestal itself, but this would always lead to plasma gradients. Finally, near critical densities can be obtained also using porous materials with a very low average density (as low as few $\textrm{mg/cm}^3$), such as aerogels  \cite{SchaeferPRL1986}, nanostructured foams  \cite{ZaniCarbon2013,ChenSciRep2016} or nanotube arrays  \cite{BinPRL2015}. \\
Using low density porous materials as a target offers some unique possibilities to control the density profile and the composition of a NCP. For example, solid foils coated with a thin (from few $\mu$m up to few tens of $\mu$m) near critical foam obtained with Pulsed Laser Deposition  have been used to study enhanced ion acceleration  \cite{PassoniPPCF2014,PrencipePPCF2016,PassoniPRAB2016}. However, while being near-critical on average, low-density materials consists in alternating voids and solid-density nanostructures. The scalelength of the density inhomogeneities might well be comparable with the laser wavelength   \cite{ZaniCarbon2013}. Modern ultra-intense laser facilities are able to provide ultra-short (down to 10s fs) pulses with an extremely high contrast, thanks to techniques such as the plasma mirror  \cite{Thaury_NatPhys2008} or the cross polarized wave generation  \cite{JullienOptLett2005}. This means that the nanostructure of the target can survive long enough to influence the interaction of the laser with the NCP. Moreover, ions would always retain their structure longer than electrons due to their significantly lower charge over mass ratio (at least as long as their dynamics is not relativistic). The nanostructure of low-density porous foams has been found to play a role even for intense nanosecond laser interaction with NCPs  \cite{GuskovPoP2011,VelechovskyPPCF2016}.\\
In the aforementioned scenarios the nanostructured nature of the target is essentially incidental: a near-critical target is desired, but such low density solids are unavoidably nanostructured. However, it is worth mentioning that laser interaction with structured plasmas with lower-than-solid density is under active investigation specifically for the physical effects allowed by the nanostructure \cite{BargstenScience2017,CristoforettiSciRep2017}.\\
In this work we investigate via an extensive two-dimensional(2D) numerical simulation campaign the role played by the nanostructure in relativistic laser interaction with NCPs. Although many structures might be worth of a detailed investigation (e.g. ordered arrays of wires), for this work we are specifically interested in random, porous nanostructures, like those typical of low-density carbon foams obtained with Pulsed Laser Deposition technique  \cite{ZaniCarbon2013,ChenSciRep2016}. \\ 
\begin{table}
	\caption{Two-dimensional simulations setup.}
	\label{tab:SimParam}       
	\begin{tabular}{ll}
			\hline\noalign{\smallskip}
		\multicolumn{2}{c}{Simulation parameters} \\
		\hline\noalign{\smallskip}
		Box size & 160$\lambda$ $\times$70$\lambda$  \\
		Points per $\lambda$ & 100$\times$100 \\
		Simulation time & 100$\lambda/c$\\
			\hline\noalign{\smallskip}
		\multicolumn{2}{c}{Plasma parameters} \\
			\hline\noalign{\smallskip}
		$n_e/n_c$ & 1, 3, 9 \\
		"\% of nanostructure" & 0\% (unf), 50\% (mix), 100\% (nano)\\
		Electrons per cell & 4 (unf), 169 (nano) \\
		Ions per cell & 2 (unf), 78 (nano)\\
			\hline\noalign{\smallskip}
		\multicolumn{2}{c}{Laser parameters} \\
			\hline\noalign{\smallskip}
		Polarization & P \\
		Angle of incidence & 0$^\circ$\\
		$a_0$ & 1, 5, 15, 45, 135 \\
		Duration FWHM & 15$c/\lambda$\\
		Waist & 5$\lambda$\\
		\noalign{\smallskip}\hline
	\end{tabular}
\end{table}
The interaction of laser pulses with uniform and nanostructured plasmas was simulated in a wide range of laser intensities. Since 2D simulations do not allow to reliably reproduce  the structure of a random porous foam, the nanostructured plasmas are modeled as a collection of high-density nanospheres \cite{CialfiPRE2016} (we note that in previous works porous media have also been modeled with collections of isolated  nano-sticks \cite{OkiharaPRE2004}). \\ We simulated three different kinds of plasmas with different initial density profiles: uniform (\emph{unf}), nanostructured (\emph{nano}) and mixed (\emph{mix}). The mixed plasmas are obtained through the superposition of a uniform plasma and a nanostructured plasma, both with halved average density. This model should simulate a partially homogenized  nanostructured plasma (e.g. due to effect of pre-heating).\\ The values of $a_0$ and $n_e/n_c$ have been chosen so that the relativistically corrected normalized electron density $\bar{n}$
\begin{equation}
\bar{n} = \frac{n_e / n_c}{\sqrt{1 + \frac{a_0^2}{2}}}
\end{equation}
 is approximately constant for some specific subsets of simulations (i.e. within the explored parameter space we can individuate families with constant $\bar{n}$). The explored range for the pulse intensity spans from the parameters of existing tabletop 10sTW facilities  \cite{NamCurrAppPhys2015} ($a_0 \sim 1$) up to those of upcoming multi-PW facilities \cite{DansonHPL2015}($a_0 > 100$), such as ELI or Apollon. \\
We note that if $a_0 >> 1$ and if the ion motion can be disregarded, the condition $\overline{n} = \textrm{const.}$ coincides with the ultra-relativistic similarity\cite{GordienkoPoP2005} $S = \frac{n_e}{n_c a_0} = \textrm{const}$ . Scaling laws of electron energy and synchrotron emission as a function of $S$ in laser interaction with highly transparent NCPs have been described by other authors\cite{HuangPRE2016}.\\
\section{Simulation setup}
\begin{figure}
	\resizebox{0.99\columnwidth}{!}{%
		\includegraphics{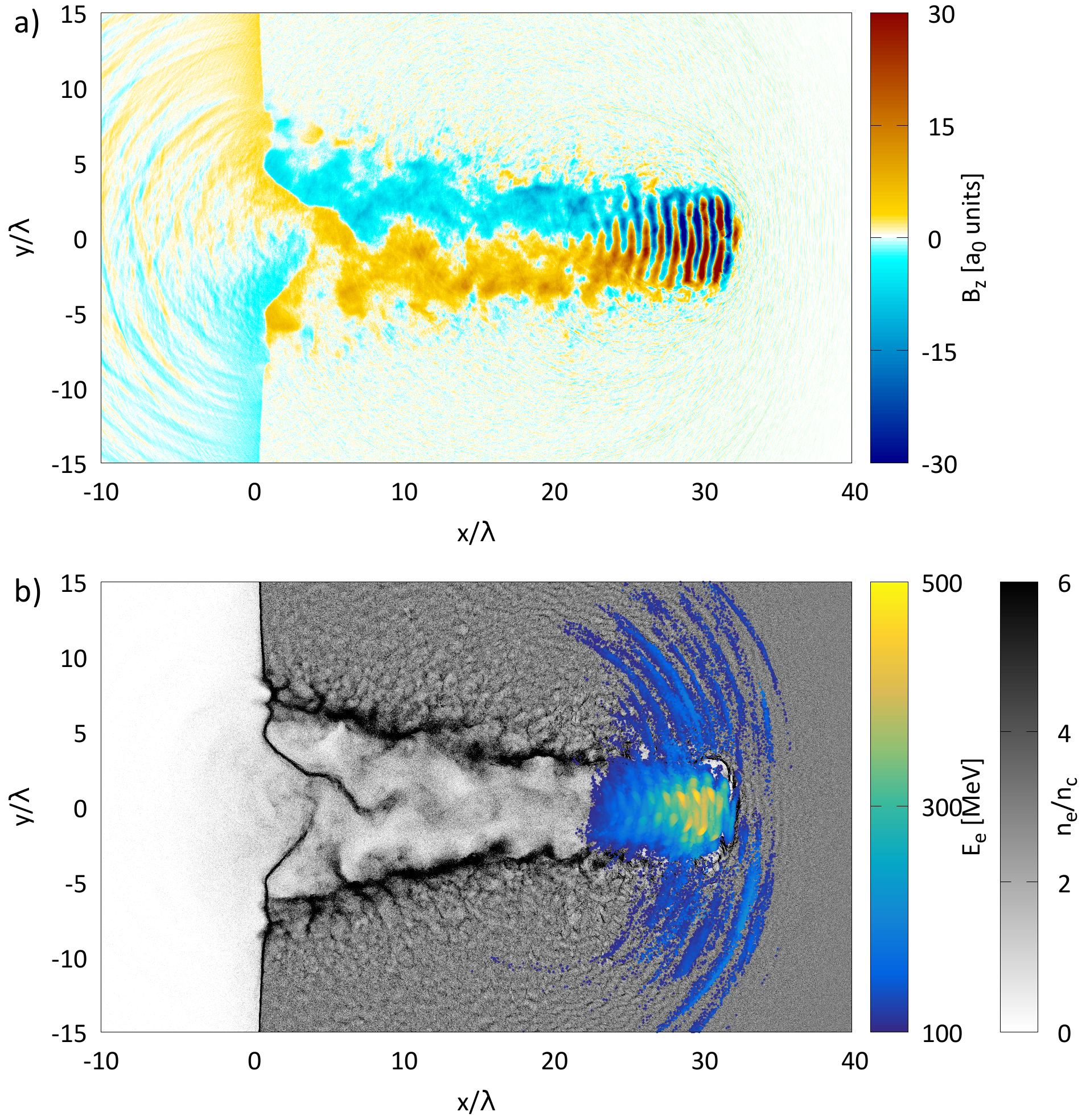}
	}
	\caption{Simulation results for the case $a_0=45, n_0=3 n_c$ (uniform plasma) at $T=50 \lambda/c$ a) Magnetic field component $B_z$ b) Electron density (greyscale) and energetic (E $>$ 100 MeV) electrons (colored dots)}
	\label{fig:ill}       
\end{figure}
The numerical simulation campaign was performed with the open source, massively parallel particle-in-cell (PIC) code \textit{piccante}  \cite{arxiv_piccante}.\\
In all simulations a laser pulse normally impinges on a 100$\lambda$-thick plasma slab, inside a box of size 160$\lambda \times$70$\lambda$ and for a simulation time of 100$\lambda/c$.  The laser pulse has a Gaussian transverse profile with a waist of 5$\lambda$ and a $\cos^2$ temporal profile with the full-width-half-maximum of the field equal to 15$\lambda/c$. The angle of incidence is $0^{\circ}$. We varied $a_0$ in the range $1 - 135$, spanning over 4 orders of magnitude in intensity. For the highest intensity values ($a_0=45,135$) additional simulations have been performed taking into account classical radiative losses using the Landau-Lifshitz model  \cite{TamburiniNJP2010,VranicCompPhysComm2016}. All NCPs are made of electrons and ions with charge-to-mass ratio equal to 2, representative of fully ionized $C^{6+}$ ions.  We let the initial average  density vary between 1$n_c$ and 9$n_c$. The density of uniform plasmas is sampled with 4 electrons per cell and 2 ions per cell. The nanostructured plasmas are modeled as collections of high-density nanospheres with 0.05$\lambda$ radius, randomly arranged in space with an overall filling factor of 0.0333\%, resulting in an electron density between 30$n_c$ and 270$n_c$. These parameters are roughly comparable with those of real carbon foams \cite{ZaniCarbon2013}, although the aim of this work is not to faithfully simulate a real material, but rather to investigate the effect of a structuring on the sub-micrometer scale. Each nanosphere is sampled with 169 electrons per cell and 78 ions per cell. All particle species are initialized with a Maxwellian distribution with a temperature of about few eV. Boundary conditions are periodic both for the fields and for the particles (the box has been chosen large enough to make recirculation effects irrelevant). The spatial resolution is 100 points per $\lambda$ along both directions, which allows to resolve the laser skin depth and to simulate plasma spheres with nanometric radii.\\
Two additional simulations were performed with a larger box (160$\lambda$ $\times$140$\lambda$, i.e. doubling the box height) and with a higher number of particles per cell (12 electrons and 4 ions per cell, uniform plasma) (not shown in the paper). We did not observe any significant difference for the time ranges of interest in this work. The main simulation parameters are reported in table \ref{tab:SimParam}. \\
Summarizing, our numerical campaign spans over the three-dimensional space of parameters ($a_0$, "percentage of nanostructure" and electron density $n_e$).
\section{Uniform near-critical plasmas}
\begin{figure*}[p]
    
		\includegraphics[width=0.95\textwidth]{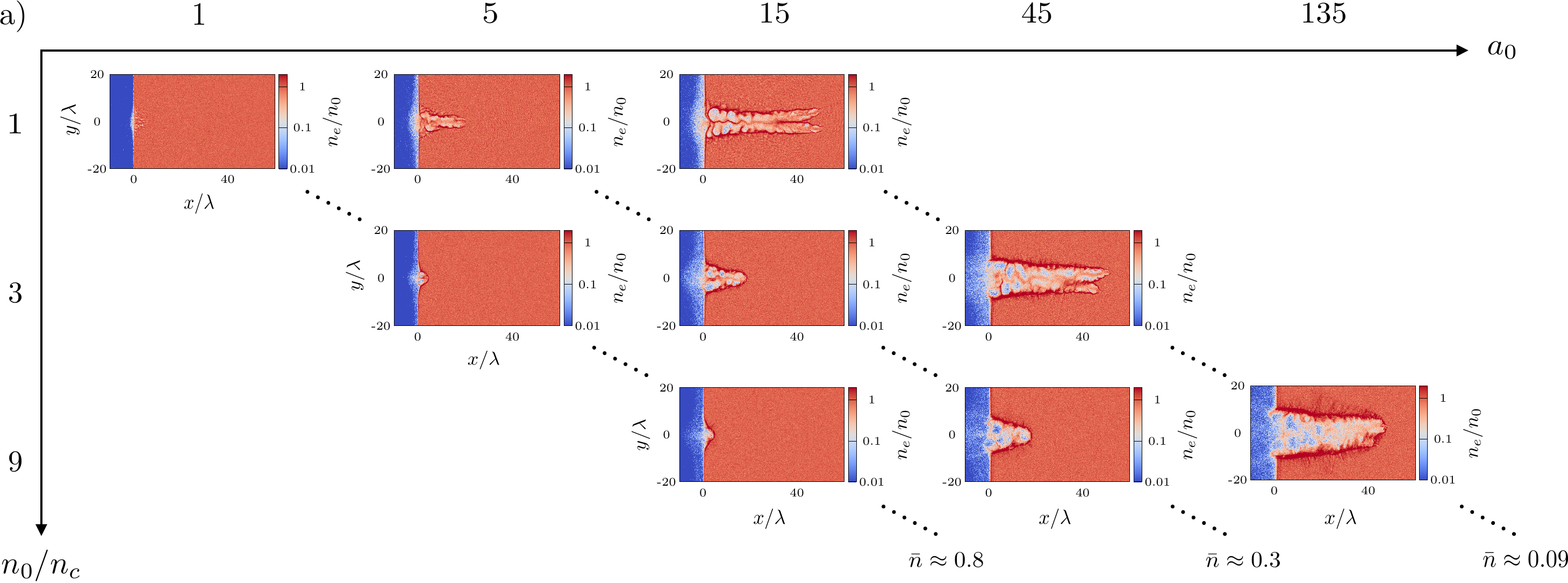}\\
	
	\vspace{1cm}	
	
		\includegraphics[width=0.95\textwidth]{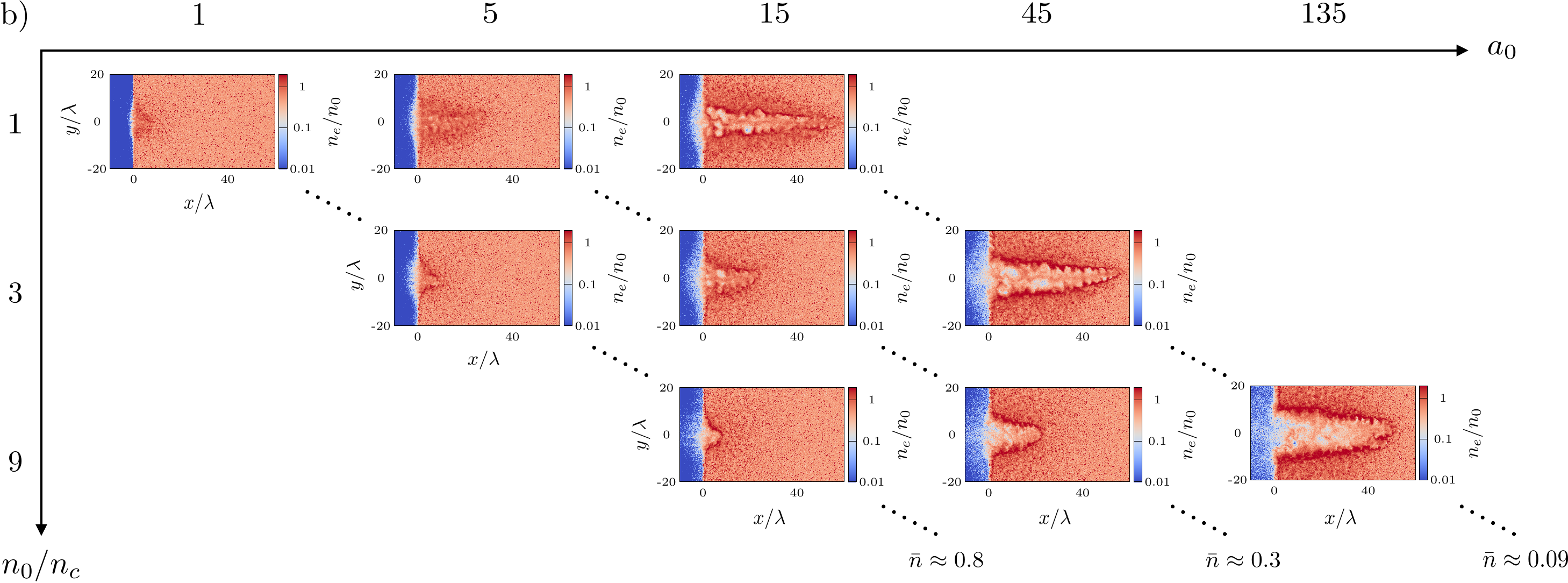}\\
	
	\vspace{1cm}		
	
		\includegraphics[width=0.95\textwidth]{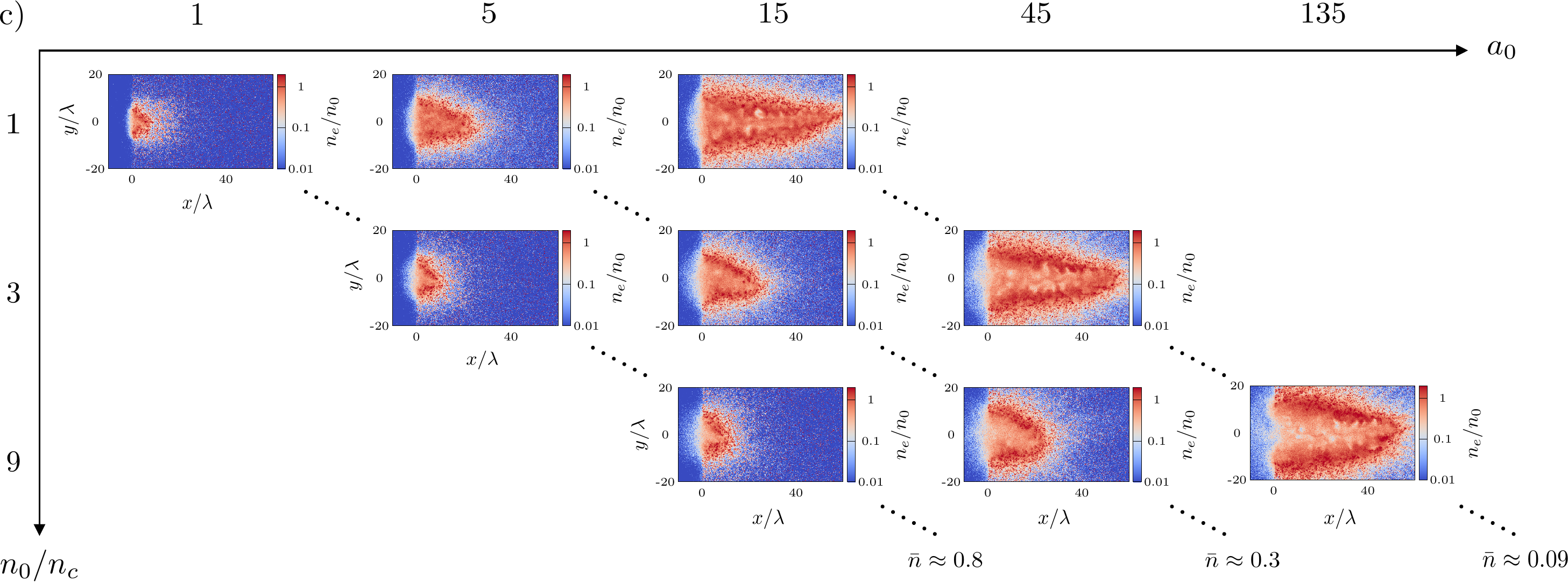}
	
	\caption{a) Electron density at time 90$\lambda/c$ for uniform plasmas. b) Electron density at time 90$\lambda/c$ for idealized nanostructured plasmas. c) Electron density at time 90$\lambda/c$ for idealized mixed plasmas. Density is normalized with respect to its initial value.
	}
	\label{fig:matrix}
\end{figure*}
\begin{figure}
\resizebox{0.95\columnwidth}{!}{
\begin{overpic}{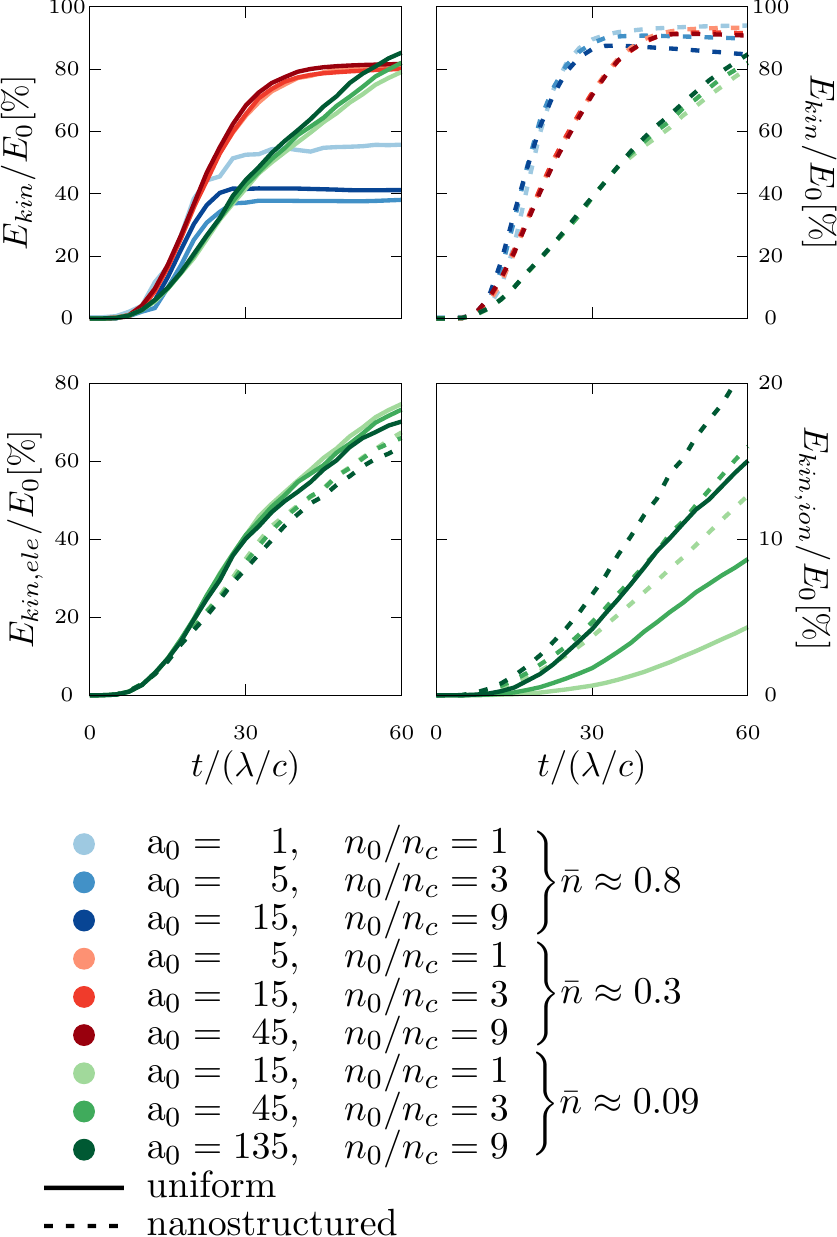}
\put(1,100){a)}
\put(65,100){b)}
\put(1,70){c)}
\put(65,70){d)}
\end{overpic}

 }
\caption{a)b) Evolution of the total kinetic energy (electrons and ions) for uniform plasmas (left) and nanostructured plasmas (right). c)d) Evolution of the electron (left) and ion (right) kinetic energy for uniform plasmas and nanostructured plasmas. Energy is normalized with respect to the total laser energy, i.e. to the total initial energy.}
\label{fig:EnergyInTime}       
\end{figure}
When an intense laser pulse interacts with a relativistically transparent plasma, it digs a channel in the electron density. Typically, although the laser pulse undergoes strong absorption, self-focusing effects take place  \cite{UmstadterScience1996,SSBulanovPoP2010}. Due to strong currents, the channel is magnetized. These general features are illustrated in figure \ref{fig:ill}, which shows a snapshot for the simulation case $a_0 = 45$, $n = 3 n_c$, uniform plasma. The upper panel shows the $\hat{z}$ component of the magnetic field (perpendicular to the simulation plane). The lower panel shows electron density in greyscale and macro electrons with energy $> $100 MeV with colored dots. A fraction of the electrons of the target undergoes a Direct Laser Acceleration (DLA) process and reaches very high energies \cite{GahnPRL1999}. \\
Figure \ref{fig:matrix} shows the electron density at $t=90\lambda/c$ for all the simulations performed for this work. In this section we will refer to panel a), which shows results for the case of a uniform plasma.  From left to right the $a_0$ parameter increases from 1 up to 135, while from top to bottom $n_e/n_c$ increases from 1 up to 9. Each diagonal is characterized by constant $\bar{n}$. In each plot the electron density is normalized with respect to the initial density.\\
On a given diagonal, the density plots show remarkable similarities. 
In the rightmost diagonal ($\bar{n} \approx 0.09$), the plasma is highly transparent and the laser is able to propagate for long distances. In all the three cases the channel reaches approximately the same point along $\hat{x}$ axis, although some differences appear in the transverse size and in the structure of the channel (e.g. in the case of $a_0 = 135$ the lateral size of the channel is about twice the size of the case $a_0=15$, and the filamentation is less pronounced). Similar features are observed in the intermediate case ($\bar{n} \approx 0.3$), where the higher plasma density prevents the laser pulse from propagating beyond $x \approx 20 \lambda$. The leftmost diagonal ($\bar{n} \approx 0.8$) includes cases where the electron density is high enough to prevent the formation of a proper channel. The physical process at play in this last group might be described as Hole-Boring\cite{WengNJP2012}, rather then laser propagation in a relativistically underdense plasma.\\
As far as the spatial distribution of the ions (not shown) is of concern, at later times ($t \gtrsim$ 70$\lambda /c$) it is very similar to the electron density, meaning that the laser pulse digs a channel in the ion density as well. However, even though ion dynamics is  always considerably slower than electron dynamics, higher intensities lead to a faster ions response. For the lowest intensity value in a given diagonal, the ponderomotive force of the laser promptly expels a large fraction of the electrons from the channel, while the density distribution of the ions remains relatively uniform for some time. On the other hand, the highest intensity cases exhibit ion density perturbations quite earlier. Indeed, ion response begins around times $70\lambda /c$, $50\lambda /c$, $30\lambda /c$ for the cases in the rightmost diagonal with the lowest, intermediate and highest intensity value respectively.\\
We note that all the simulations were performed with a sharp boundary between the vacuum and the plasma. This is clearly an idealization, since in experiments any NCP would have a transition layer (i.e. a plasma gradient from the vacuum up to maximum density). While for near-critical gas-jets or exploded targets this layer is usually very long (several $\lambda$), it can be much smaller for NCPs obtained irradiating low-density foams or similar materials (the case we are interested in this work). We expect that a gradient would not play a significant role in highly transparent cases, where the laser propagates for tens of wavelengths. However, it might well play a role for the least transparent cases. Although evaluating the effect of the gradient is beyond the scope of this work, a limited investigation of the effect was carried out. We performed few simulations of the $a_0=5$, $n_e = 3 n_c$ case, introducing a linear ramp from $0 n_c$ up to $3 n_c$ on the irradiated face. We observed a limited effect on laser penetration for a ramp of $0.5\lambda$, while penetration for significantly larger distances was observed for a $2\lambda$ ramp and for a $8\lambda$ ramp.\\
Other features observed in this simulation campaign (such as energy absorption, electron energy spectra, radiative losses) on uniform plasmas are better discussed in comparison with results obtained for the nanostructured case and are thus reported in section \ref{sec:nano}.
\section{Effect of the nanostructure}\label{sec:nano}
Figure \ref{fig:matrix}b) and \ref{fig:matrix}c) show the electron density at a given time ($t = 90 \lambda/c$) for the ``mixed'' plasma and the ``nanostructured'' plasma. Overall, these plots show strong similarities with respect to the uniform case. In particular, an analogous behavior along diagonals is observed also in these cases. As far as the electron density is of concern, the main difference between the nanostructured cases and the uniform case is that the former allows for a greater penetration of the laser pulse. Insightful features appear if we analyze the energy absorption for all the simulated cases. Figure \ref{fig:EnergyInTime}a) illustrates the temporal evolution of the total kinetic energy for uniform plasmas.
The curves for the $\bar{n} \approx 0.09$ family (green shades) are essentially superimposed as well as the curves for the $\bar{n} \approx 0.3$ family (red shades). Some differences appear within the $\bar{n} \approx 0.8$ family (blue shades), which is the least transparent case.    
For this last group the absorption efficiency is also significantly lower ($\bar{n}\approx 0.3$ and $\bar{n}\approx0.09$ families have energy absorptions $\geq 80 \%$, whereas this value drops down to $\approx 40 \%$  for the $\bar{n}\approx 0.8$ family). These results confirm the observations for the electron density: when the plasma is sufficiently transparent, keeping $\bar{n}$ constant results in a remarkably similar dynamics, even changing the pulse intensity by two orders of magnitude ($I \propto a_0^2$). Figure \ref{fig:EnergyInTime}b) shows the  temporal evolution of the same quantities for the ``nanostructured'' case (same color shades as in figure \ref{fig:EnergyInTime}a) with dashed lines). For the $\bar{n}\approx 0.09$ family the differences with the uniform foam case are negligible, while for $\bar{n}\approx 0.3$ the absorption is slightly more efficient (from $80\%$ to $\sim 90-95\%$). 
Instead, for the $\bar{n}\approx 0.8$ family, strong enhancement of absorption efficiency with respect to the uniform case is observed (the absorption efficiency is comparable with that of the other two families). Moreover the curves of this family are essentially superimposed in the ``nanostructured'' case. This is coherent with the electron densities plotted in figure \ref{fig:matrix}, which show an enhanced penetration of the laser in the ``nanostructured'' plasma, if compared with the uniform case.\\
The two lower panels of figure \ref{fig:EnergyInTime} show the evolution in time of kinetic energy absorbed by the electron population (figure \ref{fig:EnergyInTime}c)) and the ion population (figure \ref{fig:EnergyInTime}d)) for both uniform and ``nanostructured'' plasmas. Only data for $\bar{n} \approx 0.09$ family is shown. The temporal evolution of the electron kinetic energy is similar for all the simulations, with the curves for ``nanostructured'' plasmas being only slightly lower than those for uniform plasmas. More important differences appear for ion absorption. ``Nanostructured'' targets show faster ion heating, probably due to Coulomb explosion of the nanospheres. Coulomb explosion might be responsible also for the lower electron energy curves in panel c): laser energy absorption is similar for uniform plasmas and nanostructured plasmas, but in the latter case a larger fraction of this energy is promptly ceded to ions. Data for the ``mixed'' case has not been included in figure \ref{fig:EnergyInTime}, since it was always intermediate between the ``nanostructured'' case and the uniform plasma case.
\begin{figure*}
	\resizebox{0.95\textwidth}{!}{%
		\includegraphics{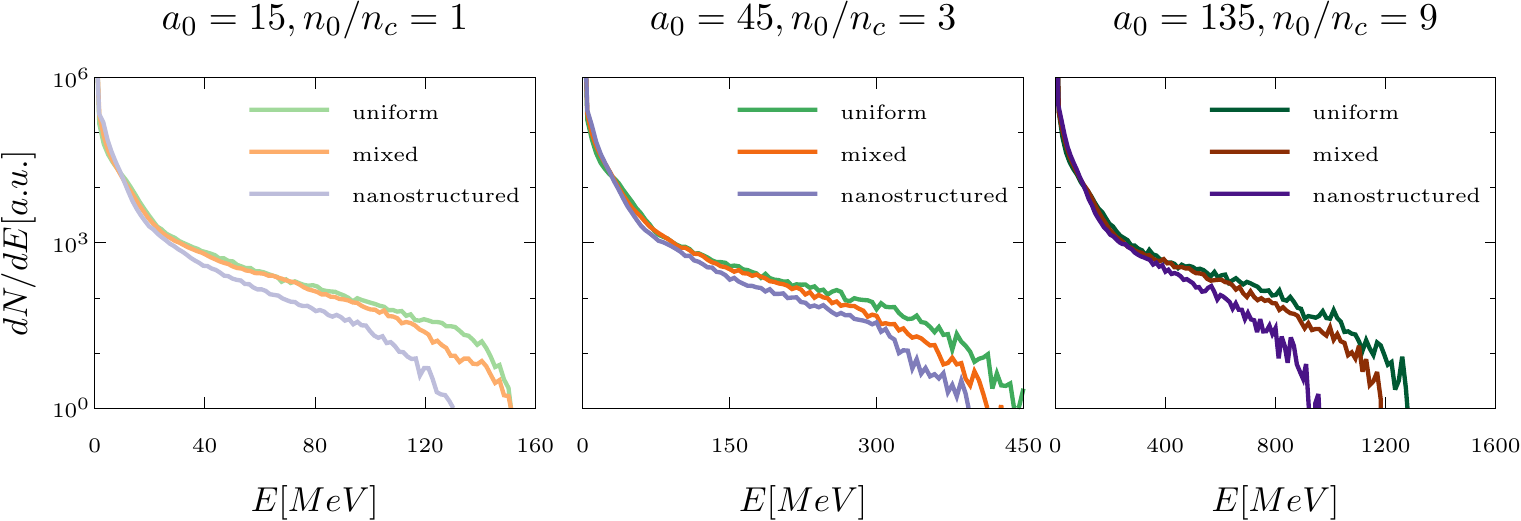}
	}
	\caption{Electron spectra at time 50$\lambda/c$ for uniform, mixed and nanostructured plasmas. It can be seen that the nanostructure leads to a suppression of the high energy portion of the spectra.}
	\label{fig:EleSpectra}       
\end{figure*}

As mentioned before, the propagation of an intense laser in a relativistically transparent plasma is associated with the forward acceleration of an electron population up to high energies. Figure \ref{fig:EleSpectra} shows electron energy spectra for the $\bar{n}\approx 0.09$ family at $t = 50 \lambda/c$ (this timestep was chosen because the acceleration process is almost finished and the electron bunch is still co-propagating with the laser). Each panel shows the spectra for the uniform plasma case, the ``nanostructured'' plasma and the ``mixed'' plasma for different values of $a_0$ within the same family. The most evident feature of the spectra is that the cut-off energy is always the lowest for the ``nanostructured'' plasma and the highest for the uniform plasma, with the ``mixed'' case in between. These simulations suggest that the presence of a nanostructure hinders the acceleration process of the high energy electron population, resulting in a suppression of the high energy tail of the spectrum.\\
\begin{figure}
\resizebox{0.95\columnwidth}{!}{
 \includegraphics{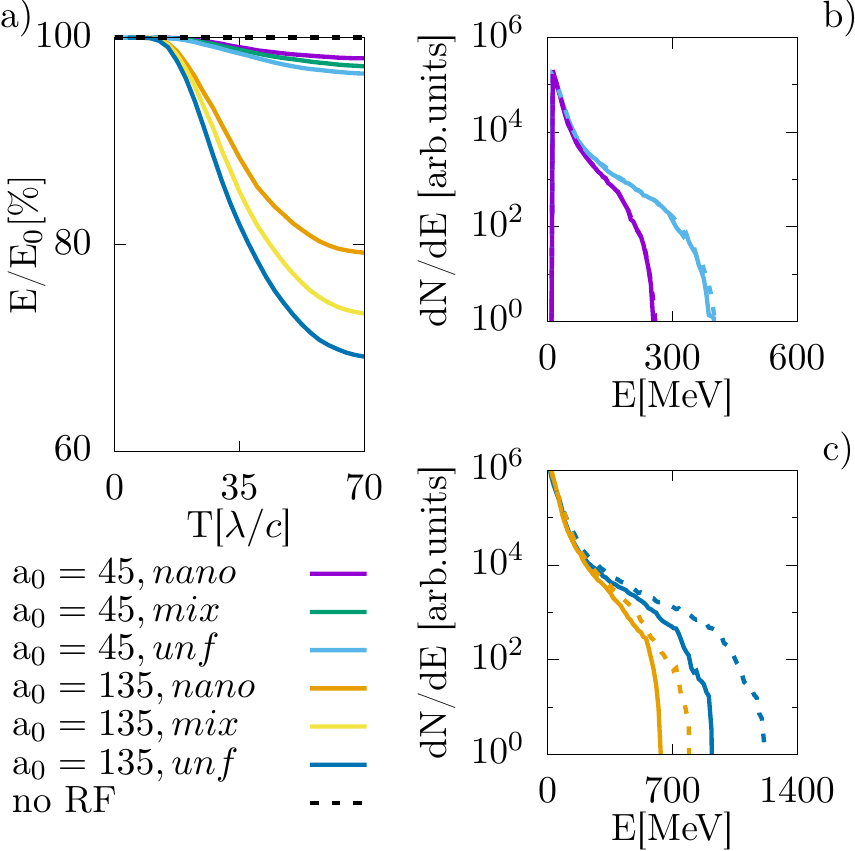}
 }
\caption{This graph shows the effect of RR for the highest intensities considered in the simulation campaign. a) Total energy as a function of time for several simulation cases with $\bar{n} \approx 0.09$. In simulations performed with $a_0 = 45$ approximately $3\%$ of the laser energy is dissipated due to RR, while with $a_0=135$ $20-30\%$ of the initial energy is lost due to RR. Simulations performed with a uniform plasma lead to energy losses about $50\%$ higher than those obtained with nanostructured plasmas. The ``mixed'' case lies in between. b) Electron energy spectra for $a_0 = 45$ (nanostructured and uniform plasma, with and without RR). The effect of RR is essentially negligible, so that the curves are essentially superimposed. Spectra were taken at $T=30 \lambda/c$ c)  Electron energy spectra for $a_0 = 135$.  Spectra were taken at $T=30 \lambda/c$.}
\label{fig:RF}       
\end{figure}
In relativistic laser-plasma interaction, charged particles undergo extreme accelerations, which are associated with EM radiation emission (essentially incoherent synchrotron emission), whose back-reaction on the particle itself is called Radiation Reaction (RR) or Radiation Friction.\\
For laser interaction with thick solid targets, radiative losses accounting for  $\sim10\%$ of the initial pulse energy can be expected for $a_0 \sim 400$ \cite{LiseykinaNJP2016}.   \\
Synchrotron emission with NCPs irradiated with ultra-intense laser pulses has been investigated by several authors\cite{NakamuraPRL2012,BradyPPCF2013,WangPoP2015,StarkPRL2016,ChangPoP2017}. In particular, high conversion efficiencies of laser energy into photon energy have been observed for highly transparent NCPs (i.e. $\overline{n} << 1$): in \cite{StarkPRL2016} for a plasma with $n_e = 4.5 n_c$ irradiated with $a_0 = 190$, a conversion efficiency $\sim 17\%$ of laser energy into synchrotron photons with energy greater than 1 MeV is obtained. For highly transparent NCPs, radiative losses are mainly due to the motion of highly relativistic electrons (co-propagating with the laser pulse) in the EM field resulting from the combination of the laser field and of the self-generated magnetic field in the channel \cite{StarkPRL2016,Chang2017SciRep,ChangPoP2017}. According to these works, it is reasonable to expect high radiative losses for the $a_0 = 135,~ \overline{n}= 0.09$ cases in our simulations. Thus these cases have been re-run enabling classical RR\cite{TamburiniNJP2010,VranicCompPhysComm2016}. Simulations with RR have been run also for the $a_0 = 45,~ \overline{n}= 0.09$ cases, since radiative losses $\sim 3\%$ have been reported for similar intensities and tailored plasma density profiles \cite{NakamuraPRL2012} (i.e. an exponential ramp from the vacuum up to the solid density).  \\
The leading term of RR force
\begin{equation}
F_{RR} \approx - \left({\frac{2}{3} \frac{r_e}{\lambda}}\right) \gamma^2 \left[ {f_L^2 - (\mathbf{v}\cdot\mathbf{E})^2} \right]\cdot \mathbf{v} + \ldots
\end{equation}
depends quadratically on $\gamma$ ($r_e$ is the classical electron radius, $f_L$ is the Lorentz force, $\mathbf{v}$ is the particle velocity normalized with respect to $c$, $\mathbf{E}$ is the electric field normalized with respect to $e \lambda / m_e c^2$).\\
Figure \ref{fig:RF}a) shows RR effects on total energy for few simulation cases with $\bar{n} \approx 0.09$. Without RR, the energy is conserved by the code with an accuracy $\lesssim 2$\textperthousand, thus losses greater than $\sim 1\%$ can be attributed to radiation emission. As expected, radiative losses are modest (few $\%$) for $a_0 = 45$, while for $a_0 = 135$ up to one third of the initial pulse energy is lost at the end of the simulation. In this last case, differences between the three plasma types are particularly evident: conversion efficiency of laser energy into synchrotron radiation attains about $20\%$ for the ``nanostructured'' plasma and about $30\%$ for the uniform plasma (the ``mixed'' case lies between these two extremes).
These results are reasonable considering the strong dependence of radiative losses on particle energy and the fact that in simulations performed without RR a suppression of the high-energy tail of the electron energy distribution was observed for the ``mixed'' and the nanostructured cases (see figure \ref{fig:EleSpectra})).
 Since also the spectrum of the emitted radiation (not simulated by our PIC code) depends crucially on $\gamma$ of the emitting particles, it is reasonable to expect significant differences between the considered cases.\\
Figure \ref{fig:RF}b)c) show the effect of RR on electron energy spectra. As expected, the effect is negligible for the case $a_0 = 45$, whereas RR strongly affects the high energy tail of the spectra for $a_0=135$.
\section{Conclusions}
In this work we have presented an extensive 2D numerical campaign to assess the effect of a nanostructure on laser interaction with near-critical plasmas. The nanostructure (modeled as a collection of spheres) was found not to dramatically alter the qualitative features of laser interaction with near-critical plasmas (i.e. the formation of a magnetized channel). However, a significant effect was found on electron energy spectra and conversion efficiency of laser energy into ion kinetic energy. In particular, a suppression of the high energy tail of electron energy spectra was observed for nanostructured plasmas in comparison with uniform plasmas, while the nanostructure was found to allow for higher conversion efficiencies of laser energy into ion kinetic energy. Finally, for the highest simulated intensities ($a_0 = 135$), a significant  effect of the nanostructure on radiative losses was observed.\\
Our work might suggest possible strategies for an experimental investigation of the effect of the nanostructure in laser interaction with near-critical plasmas obtained from low-density foams or similar materials. In an actual experiment, a controllable pre-pulse  \cite{KahalyPRL2013} or other forms of pre-heating \cite{ChenSciRep2016} could be used to initiate a homogenization of the target density. This would allow to tune the uniformity of the target before the arrival of the main pulse, simply changing its delay with respect to the pre-heating.
As an example, observables such as the energy spectrum of the electrons escaping from the target could be collected as a function of this delay (i.e. for different states of homogenization of the target).
\section*{Acknlowlegdments}
The research leading to these results has received funding from the European Research Council Consolidator Grant ENSURE (ERC-2014-CoG No. 647554). We also acknowledge LISA access scheme to MARCONI HPC machine at CINECA(Italy) via the projects SNAP and EneDaG.

\section*{Author contribution statement}
A.F and L.F. performed the numerical simulations. M.P. and C.E.B. supervised the work. The paper was mainly written by L.F. with the assistance of A.F. All authors reviewed the manuscript. A.F. designed the graphical abstract.

\bibliographystyle{ieeetr}

\end{document}